\documentclass[11pt,twoside]{article}
\usepackage[hmarginratio=1:1,top=32mm,columnsep=20pt]{geometry}
\usepackage[footnotesize,labelfont=bf,up,textfont=it,up]{caption}
\usepackage{float}
\usepackage{booktabs}
\usepackage{paralist}
\usepackage{amsmath}
\usepackage{hanging}
\usepackage{wasysym}
\usepackage{graphicx}

\usepackage{abstract}

\usepackage{titlesec}
\renewcommand\thesection{\Roman{section}}
\renewcommand\thesubsection{\Roman{subsection}}
\titleformat{\section}[block]{\large\scshape\centering}{\thesection.}{1em}{}
\titleformat{\subsection}[block]{\large}{\thesubsection.}{1em}{}

\usepackage{fancyhdr}
\title{\textsc{A Reputation-Based Model for Decision-Making in Online Social Networks}}
\author{Stan Palasek\\Department of Mathematics, Princeton University}
\date{September 2014}
\begin{document}

\maketitle

\thispagestyle{fancy}

\begin{abstract}
\vspace{-.1in}
The online exchange of social recognition including, for instance, the Facebook ``like'' appears to produce a scarce allocation without a clear utility function defined for anyone involved. Given the importance attached to such digital commodities by both users and advertisers, it is of interest to study the forces governing their economics. Here we propose a centrality measure akin to eigenvector centrality to describe an individual's perceived importance in an online social network. It is shown \textit{in silico} that strategically maximizing this prestige metric results in finite nontrivial rates of ``like'' endowment. Furthermore, it is found that systems of reputation-seeking agents are supported most robustly by networks with the features of real human societies including preferential attachment and the small-world property. We conclude that the incentive system studied here can produce realistic behavior and may therefore provide a framework for a more general model of decision-making in online communities.
\end{abstract}

\section{Introduction}
The rise of the Internet as a means of exchange for products and ideas raises economic and legal challenges unlike those faced before (Castells 2000, pp.~471-475). One feature of the ``Web 2.0'' economy is the new ubiquity of goods that may have significant costs of production but can be distributed at zero or negligible marginal cost including, for instance, digital media, software, and access to online services. In lieu of a unified infrastructure to allocate such goods and prevent illegal sharing (though several frameworks have been proposed to date; see Aigrain 2012 and Fournier 2014), two distinct regimes for the distribution of digital commodities have emerged. The first and most conventional is strictly profit-based, mandating that consumers pay producers for content and services either directly through payment or indirectly by being subjected to advertising (Fuchs 2009). Alternately and bafflingly from a traditional economic perspective, some creators opt to distribute what they have produced without explicitly requiring anything in return. Although this phenomenon is found throughout the web, it is perhaps best-documented as it applies to open-source software (Lerner \& Tirole 2002) and digital information more generally (Weber 2000; Curien \& Muet 2004). One might reasonably ask what motivates developers to place on the public domain their potentially lucrative software creations, the likes of which include products as universal as Apache and Linux. It has been suggested that online ``gift economies'' are impelled by an expectation of reciprocity by all members (Veale 2003), effectively enforcing a norm of reciprocal altruism which has indeed been shown to be stable both in nature (Trivers 1971) and in particular online communities (Cohen 2003). Another hypothesis attempting to explain the unexpected degree of online generosity is one of reputation, the pursuit of which has been established as a powerful social force (Appiah 2011). Eric Raymond, a prominent proponent of the ``open-source movement,'' notes that ``the `utility function' Linux hackers are maximizing is not classically economic, but is the intangible of their own ego satisfaction and reputation among other hackers.'' (Raymond 2001, p.~53)

Raymond's is the model by which we will proceed in considering the exchange of yet another intangible product without marginal cost: gratification constructs in online social media, most significantly the Facebook ``like.'' The philosophical and economic exchanges of such commodities have been described in detail in the context of the corporate accumulation of social favor as a means of promotion (Arvidsson 2009; Gerlitz \& Helmond 2011). ``Likes'' for advertising purposes, unlike goods such as open-source software, can be valuated quantitatively as they are literally traded on a market (Bilton 2014). However, in such markets, supply and demand are governed by the availability and costs associated with armies of ``bots'' and the promotional needs of the parties who are purchasing the online recognition. The acquisition of these ``likes'' thus becomes a matter of technology and marketing rather than of the social forces to which we will direct our attention in this work. The lack of a monetary incentive when communities of ``friends'' decide whether to positively recognize the content of one another may shed light on the factors that drive this and other gift economies.

\section{Defining Social Prestige}

In order to examine why ``likes'' are traded within social networks, one must first determine the motivations of people in using Facebook and other social networks at all. Central purposes identified in the literature include, among other things, maintaining social contacts, surveilling acquaintances (Joinson 2008), increasing one's credibility (Jessen \& J{\o}rgensen 2011), and improving one's social standing (Ellison et al.\ 2007). We will consider only the last of these factors in this work because it is most likely to produce economic results; namely, while ``likes'' are not scarce, social capital is. It is therefore reasonable to consider a system in which agents act strategically to acquire it. To do so we must develop a method of quantifying an agent's popularity or prestige within a network in which ``likes'' are exchanged. Certainly the problem of identifying important nodes in a network has been considered at length in the literature, producing metrics to address a wide variety of questions including those of simple geometric features (degree centrality), each node's nearness to the rest of the graph (closeness and betweenness centralities) (Dehmer 2011, pp.~12-13), and the diffusion of information and ideas (Banerjee et al.\ 2013), among others. A key requirement for the centrality measure to be defined here, to which we shall refer as ``likedness centrality,'' is that to be ``liked'' by a prominent member of society allows the recipient to share in the granter's social capital. This feature is reminiscent of both eigenvector centrality and Google's PageRank algorithm (Brin \& Page 1998). However, we will see in the next section that these models are inappropriate as they neglect the potential for economic inflation of gratification constructs.

Consider a social network of $N$ nodes labeled $\{1,2,\ldots,N\}$ which are arrange in an unweighted and undirected graph $G$.\footnote{We focus on an undirected, unweighted graph with Facebook's network in mind. However, one may easily introduce asymmetry and arbitrary real values to the adjacency matrix to respectively permit directed connections (eg.~Twitter, Instagram, citation networks) and idiosyncratic degrees of interaction.} The graph is equipped with an $N\times N$ matrix $R$ of non-negative reals with the property that if an edge $(i,j)\notin G$, then $R_{i,j}=R_{j,i}=0$.\footnote{The converse need not hold because we do not require that adjacent nodes exchange nonzero ``like'' rates in both directions.} We shall interpret the entry $R_{i,j}$ as the rate at which the $j$th agent ``likes'' content from the $i$th. For fixed $G$, we seek to define the $N$-vector $\mathbf{L}(R)$ whose $i$th coordinate contains the prominence or ``likedness centrality'' of the $i$th agent. We might proceed in parallel with eigenvector centrality and define $\mathbf{L}(R)$ to be the vector satisfying
\begin{align}
L_i(R)=\frac{1}{\lambda}\sum_{j=1}^NR_{i,j}L_j(R)
\end{align}
for the $\lambda$ of largest possible magnitude. This formulation is problematic. Each agent observes that as he ``likes'' more content, the recipients gain prestive, thus inflating the value of whatever ``likes'' he receives in return. Since the explicit cost of ``liking'' content is negligible, this would imply that any pair of rates that are not both initially zero will increase \textit{ad infinitum}, provided that agents are rational.

The fundamental issue with using this or another established centrality measure is that they compute prominence from the standpoint of an outside observer of the network. They are therefore unsuited to model the view from an agent's perspective, off of which his utility function must be based. We attempt to resolve this by allowing each individual to judge his prestige in the network relative to the prestige of others whom he sees; namely
\begin{align}
L_i(R)=\sum_{j=1}^NR_{i,j}L_j(R)\bigg\slash\sum_{j=1}^NG_{i,j}L_j(R)
\end{align}
where $G_{i,j}$ is the entry at $i,j$ in $G$'s adjacency matrix.\footnote{Scaling both $\mathbf{L}(R)$ and $R$ (effectively changing the unit in which the rates are measured) does not affect the equation. Although the key results of this paper depend only on relative values of $\mathbf{L}$, for the sake of uniqueness when stating absolute values we adopt the convention of letting $\mathbf{L}(R)\cdot\mathbf{1}=1$.}

In this model, when an agent increases the rate at which he delivers ``likes,'' two opposite effects must be considered. First, as before, when the recipients' prestige increases so does the value of the ``likes'' with which they reciprocate. Second, as the total prestige of the network's members grows, each agent's own prominence appears lower to him or herself by comparison. Each member of the network must strategically balance these two competing forces, perhaps yielding nontrivial equilibrium rates of ``like'' exchange.

\section{Tit-for-tat Phenomena}

One potential model for agents' decision processes involves the reciprocity arrangement suggested for the broader internet economy by Veale (2003). It might be supposed that there exists a coalition of agents who can together produce a desirable outcome for the group but individually have incentive to defect. With a distant time horizon, it has been established that it may be nonetheless rational for individuals to coordinate their actions, producing what may resemble \textit{quid pro quo} ``like'' exchanges (Axelrod 2006, p.~31). In this work we will avoid this complication by assuming throughout that the discount factor is large, due to either the forgetfulness of the players or the transience of online social networks. Otherwise, this problem would resemble a repeated multiplayer prisoner's dilemma with a highly complex payoff matrix. As even the apparently simple two-player repeated prisoner's dilemma lacks an infallible best response, such a problem would be intractable.

\begin{figure}
\begin{centering}
\includegraphics[scale=.5]{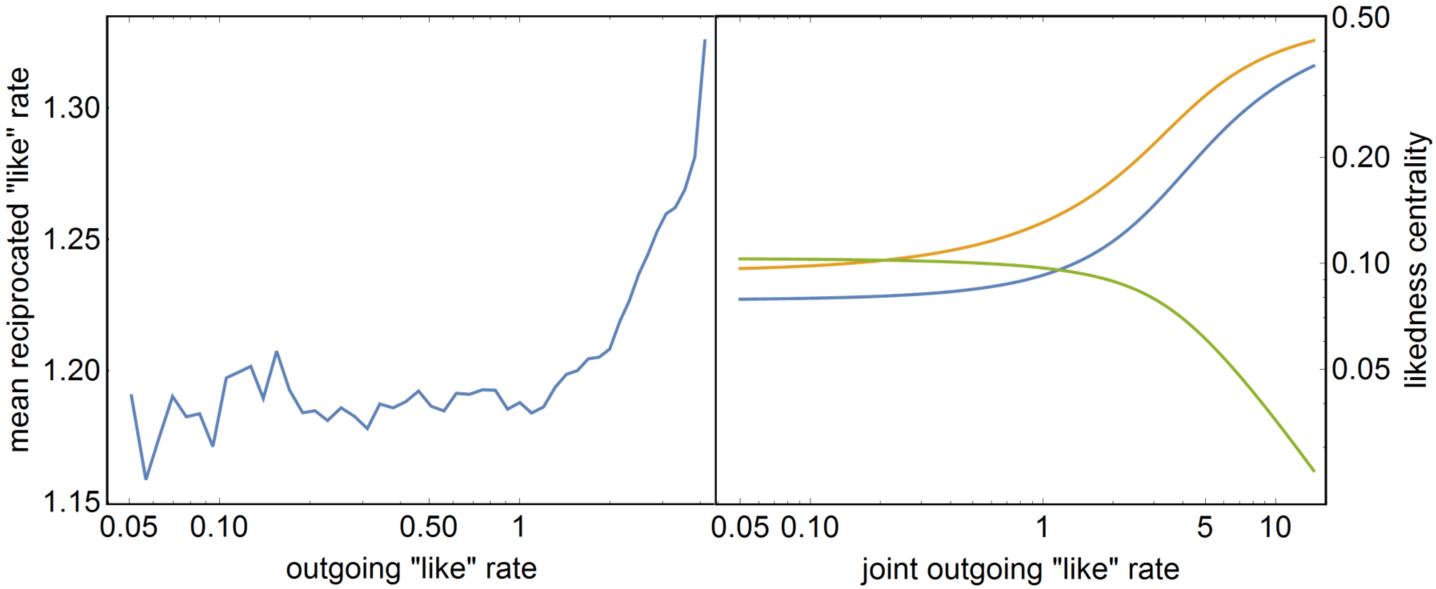}
\caption{At left is the mean ``like'' rate received in return for various outgoing rates exhibited by the most strategic of the $10^6$ societies described in the next section. We then allow two outlying nodes to form a coalition and ``like'' one another at a joint rate. The right pane shows the effect on the likedness centralities of the coordinating agents (blue and orange) and the rest of the network (green).}
\end{centering}
\end{figure}

Even when players have no concern for the prospect of future cooperation, the proposed likedness centrality model is intrinsically able to encourage tit-for-tat play in the following manner. If an agent is to be strategic about whom he ``likes'' (which he must, as ``liking'' in excess diminishes his own prestige in comparison to his peers'), then he must ``like'' those who would deliver the greatest payoff to him: namely, those who already ``like'' him more frequently, for he gains directly from their inflated prominence. Thus reciprocity in moderation becomes a self-enforcing behavior favorable to both parties. However, unlike the eigenvector model, inflation occurs as the network is flooded with ``likes'' and those involved in tit-for-tat exchanges begin to experience decreasing marginal returns. This theory is robustly apparent in the simulated behavior of pairs of rational agents, shown in the first frame of Figure 1. The second frame shows the inevitable corollary that when cooperative strategies are allowed, it may be rational for coalitions of agents to ``like'' one another \textit{ad infinitum}.\footnote{We will not overly concern ourselves with this consequence of the short time horizon assumption because in any real community, the rest of the network would likely recognize the \textit{quid pro quo} arrangement and the coalition's credibility would suffer.}

\section{Simulating Strategic Social Agents}

The model society in this investigation will be a social network randomly generated by a Barab\'{a}si-Albert graph distribution with $N=10$ and $k=2$ which is known to exhibit realistic structural features including preferential attachment, small-world dynamics, and clustering to a greater extent than a random graph (Barab\'{a}si \& Albert 1999). Even with the simplification of a short time horizon as assumed in the last section, finding equilibrium rate matrices is a matter of solving a system of $2|G|=34$ nonlinear differential equations in as many variables. Although such a calculation might be tractable numerically on a small number of graphs, to carry it out over a representative subset of all possible communities is not. Instead, for each of $10^6$ Barabási-Albert graphs, we generate a rate matrix with entries selected independently and at random according to an exponential distribution with $\lambda=1$. This allows us to rank the pairs $(R,G)$ on a spectrum of stabilities based on the response of $L_i (R)$ to perturbations of the $i$th column of $R$, ie.\ the degree to which $i$ is motivated to change the rates at which it endows ``likes.'' In particular, define the stability of a system as
\begin{align}
S(R,G)=\exp\left[-\sum_{(i,j)\in G}\left(\frac{\partial L_i(R)}{\partial R_{j,i}}\right)^2\right]
\end{align}
where the sum separately counts both directions of each edge.\footnote{Analytic methods fail to find explicit forms for likedness centrality or its gradient in all but in the simplest of systems. The derivatives required in the stability formula are instead computed as finite difference quotients by perturbing entries of the rate matrix by 1\% of their values.} When no individual has incentive to defect, the appropriate partial derivatives vanish and the system has unitary stability. This is the metric we will use henceforth to evaluate whether the matrix $R$ can be produced by rational agents over a graph $G$.

\section{Rationality of Finite ``Liking''}

A central question is whether it is strategic for agents to deliver ``likes'' at all, and if so whether the model presented here correctly predicts that allocation occurs at finite rates in equilibrium. We will consider the $0.1\%$ most stable of the $10^6$ systems defined in the previous section, corresponding to $S(R,G)\leq0.00125$, to be strategic. Figure 2 illustrates how over- or under-represented each strategic ``like'' rate interval is compared to the random exponential population whence it is drawn. It is seen that while strategic agents are unlikely to exhibit very low rates, they prevail among those who ``like'' at rates past the $1-e^{-1}\approx63.2$ percentile. An extremum occurs at the $95.3$ percentile rate which is populated at a $52.9\%$ greater frequency by individuals acting strategically. Agents with rates past this peak increasingly populate unstable systems, and indeed at the $99.2$ percentile non-strategic agents prevail. We can therefore conclude that the systematic pursuit of likedness centrality drives rational individuals to a finite but nonzero rate of bestowing ``likes,'' as would be expected in any plausible online social network.

\begin{figure}
\begin{centering}
\includegraphics[scale=.7]{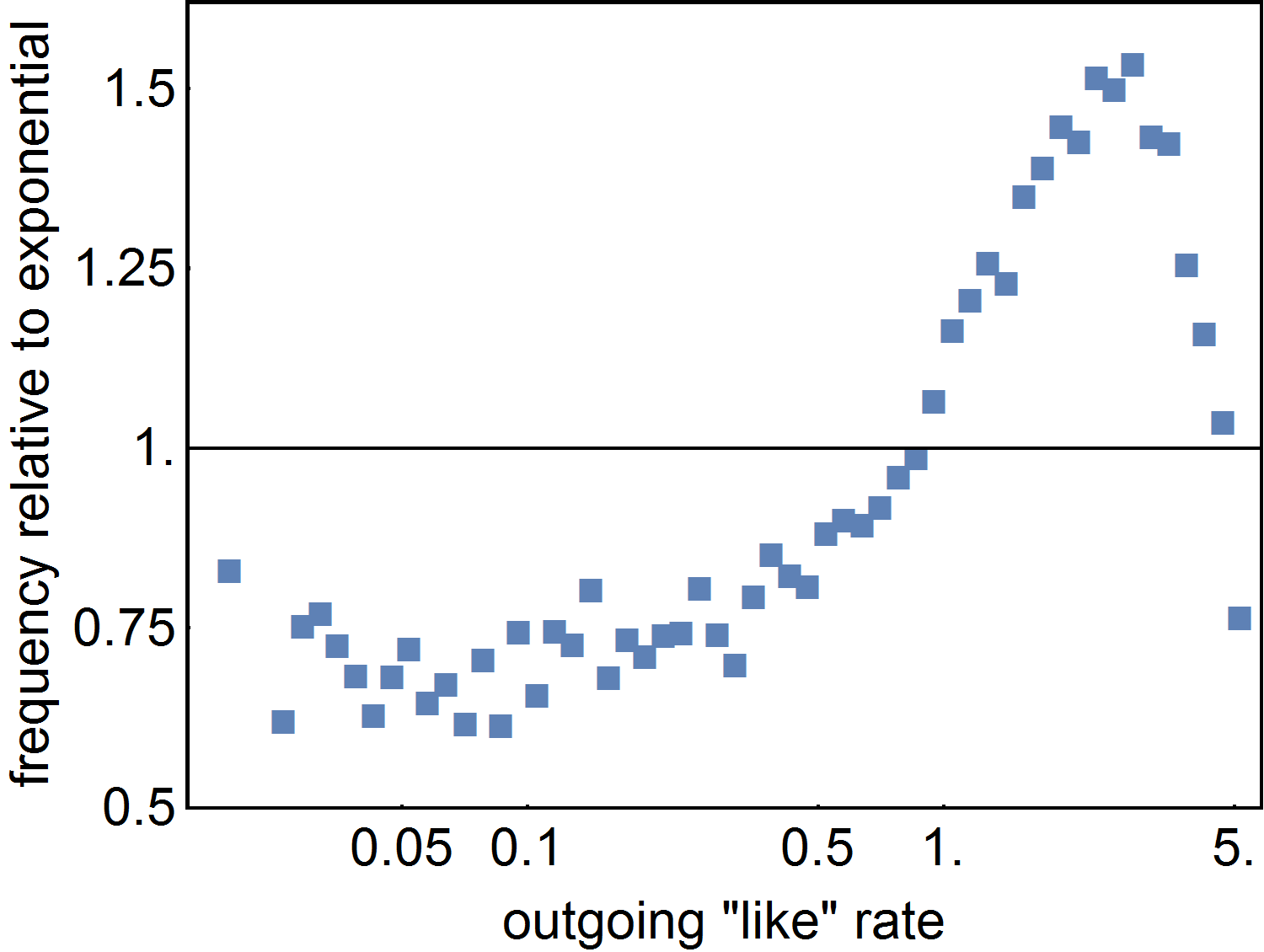}
\caption{Frequencies of various outgoing ``like'' rates for strategic agents relative to the random exponential model.}
\end{centering}
\end{figure}

\section{Emergent Network Features}

Because each of the sample societies considered in the simulation is arranged in an independently-generated graph, the same dataset allows us to investigate what network features are most likely to come about when the ``like'' exchange ensemble is in equilibrium. We will examine three properties which are characteristic of realistic networks: a long-tail degree distribution, low mean graph distance, and high clustering (Albert \& Barab\'{a}si 2002). All three are, to various extents, already present in the Barab\'{a}si-Albert model so we wish only to determine if strategic selection accentuates them further.

\begin{figure}
\begin{centering}
\includegraphics[scale=.7]{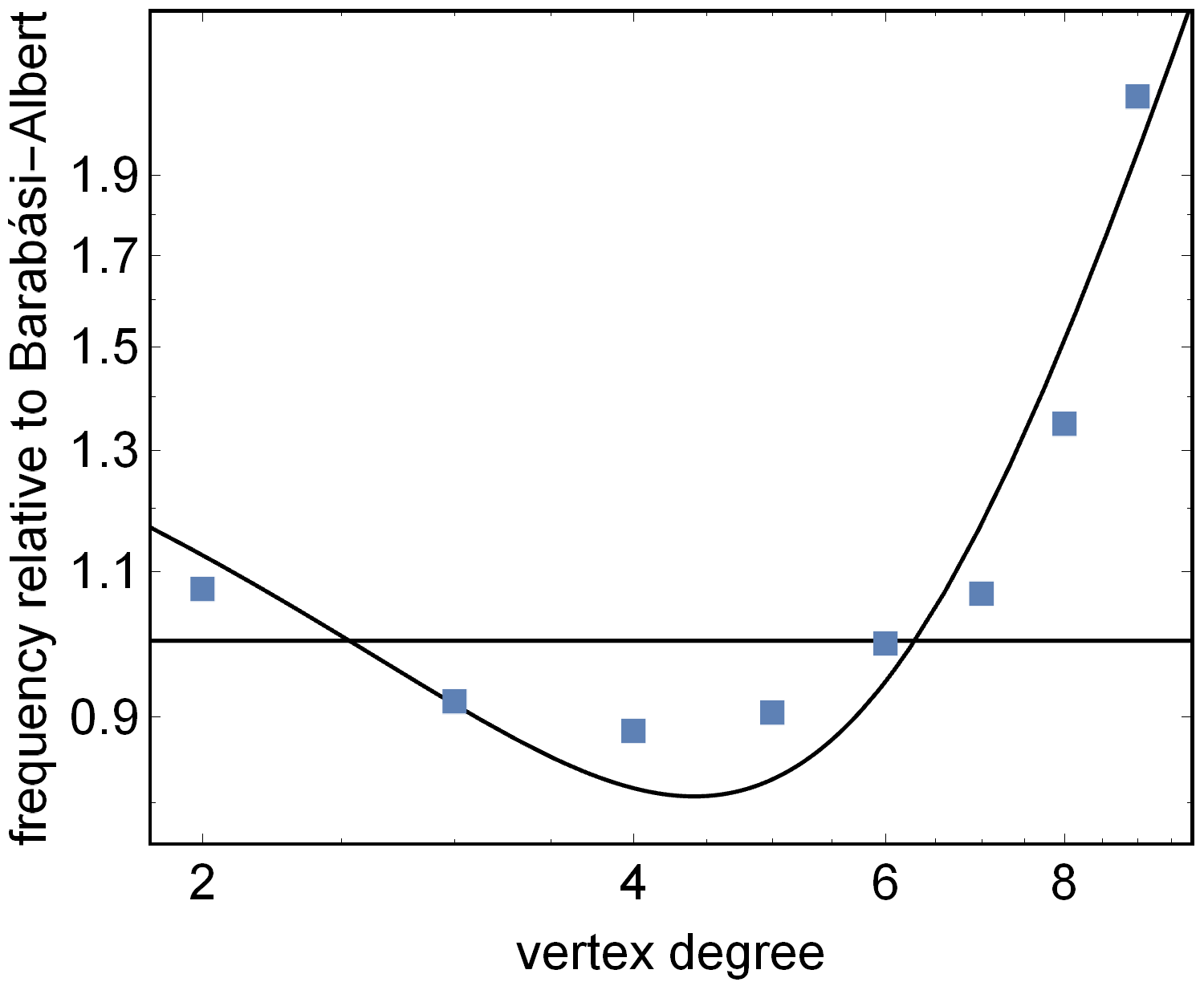}
\caption{The frequency of each possible vertex degree for strategic societies compared to the baseline Barab\'{a}si-Albert model. The horizontal line shows where the models would equate.}
\end{centering}
\end{figure}

To examine the relative level of preferential attachment in strategic networks, we use the same class of the $0.1\%$ most stable systems from the last section. We then compute the degree distributions for both this subset and the random population as a whole. The frequency of each degree is compared between the two populations in Figure 3. It is seen that the overrepresentation of the spectrum of degrees is ``U''-shaped, with both its tails occurring more frequently in strategic systems. Dramatically, fully connected nodes (of degree $N-1=9$) are $2.11$ times as common in strategic systems. Despite the intrinsic scale-free nature of Barab\'{a}si-Albert networks (for they are constructed using explicit preferential attachment), an inequitable degree distribution is still more prevalent among graphs which support equilibrium ``like'' rate matrices. Next we will direct our focus toward the so-called ``small-world'' property which will be measured as follows. For any distinct pair of vertices labeled $(i,j)$, let $\ell_{i,j}$ be the minimal length required to travel from $i$ to $j$. The mean path length is the average of these geodesic distances which is charactersitically small (typically on the order of $\log N$) for small-world networks (Watts \& Strogatz 1998). Figure 4 depicts the dependence of a system's mean stability on the various possible mean path lengths (for there is a discrete spectrum of the latter). A robust negative correlation is found, implying that low-diameter networks are better suited on average to support equilibrium ``like'' ensembles.

\begin{figure}
\begin{centering}
\includegraphics[scale=.7]{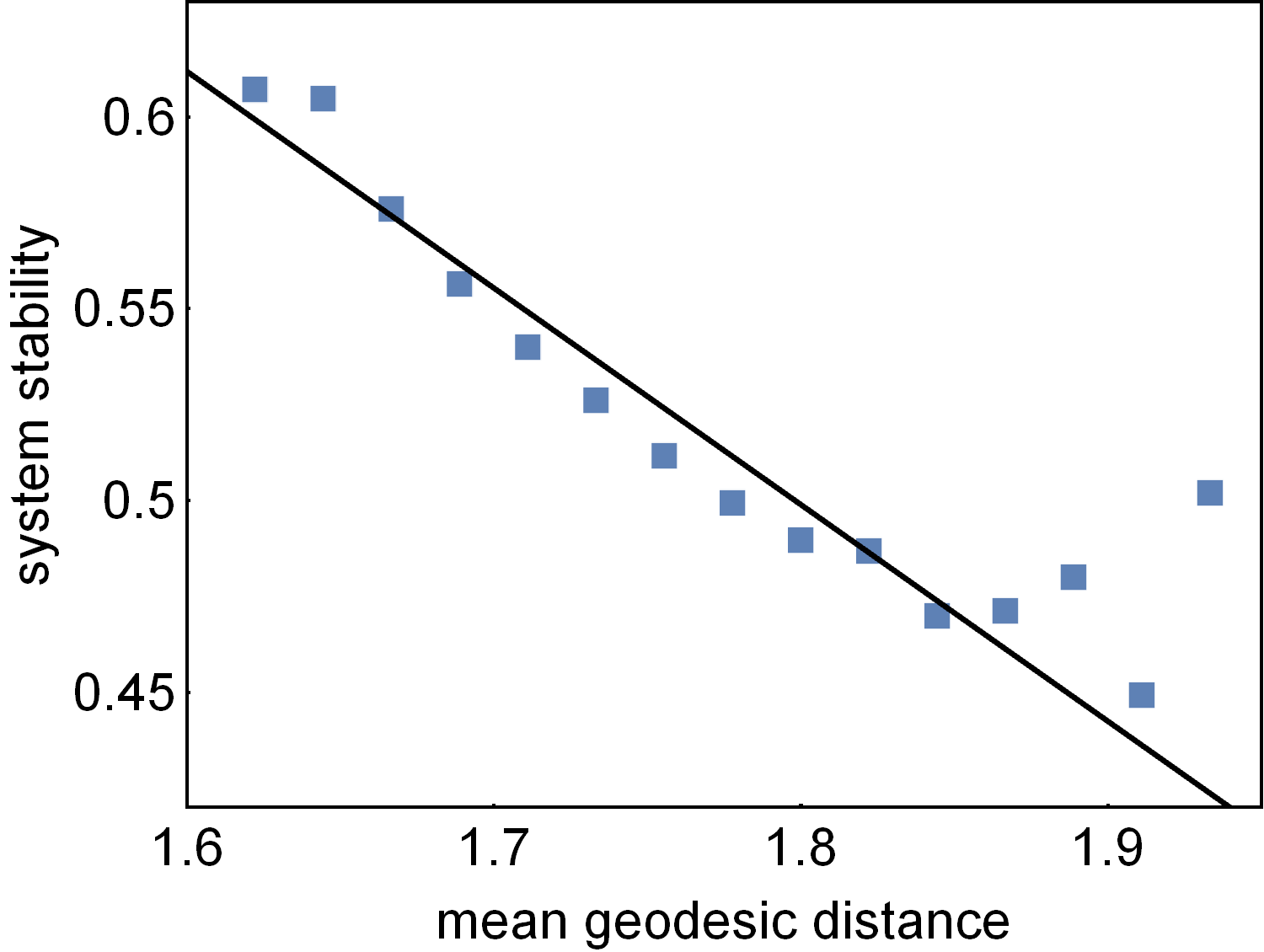}
\caption{Effect of the graph's mean degree of separation on the stability of the $(R,G)$ system.}
\end{centering}
\end{figure}

The final structural network feature to be examined is clustering, ie.\ the tendency of those who share mutual friends to be adjacent. Mean local clustering is defined as follows in the manner of Watts and Strogatz. For a given vertex labeled $i$, let the local clustering coefficient be the proportion of potential connections between its ``friends'' which are in fact edges of the graph. The mean clustering is the mean of these local coefficients across all nodes $i$. As in the previous section, we plot the effect of the clustering coefficient on the stability of the system, shown in Figure 5. We find that the resulting scatter plot is in fact partitioned into approximately eight near-exponential series, perhaps corresponding to some discrete structural feature of the network. The overall negative trend indicates that clustering does not tend to promote strategic ``like'' rate equilibria.

It is not lost on the author that these three global metrics are mutually dependent, particularly when computed for such small communities. It is therefore prudent to determine which of the identified relations are truly due to variation in the explanatory variable and which exist only because the explanatory variable correlates strongly with another factor. To this end we regress the system's stability to a logistic function against the three variables examined in this section using a damped least-squares method.\footnote{Preferential attachment is quantified as the standard deviation of the graph's degree distribution and the other variables are defined as before.} The fitted coefficients for preferential attachment, mean path length, and mean local clustering are respectively $0.0973$, $-0.2757$, and $-0.0113$. All three trends have signs as observed in isolation so we need not qualitatively modify our conclusion. However, the effect of clustering can be almost entirely accounted for by variation in the presence of the scale-free and small-world properties.

\begin{figure}
\begin{centering}
\includegraphics[scale=.7]{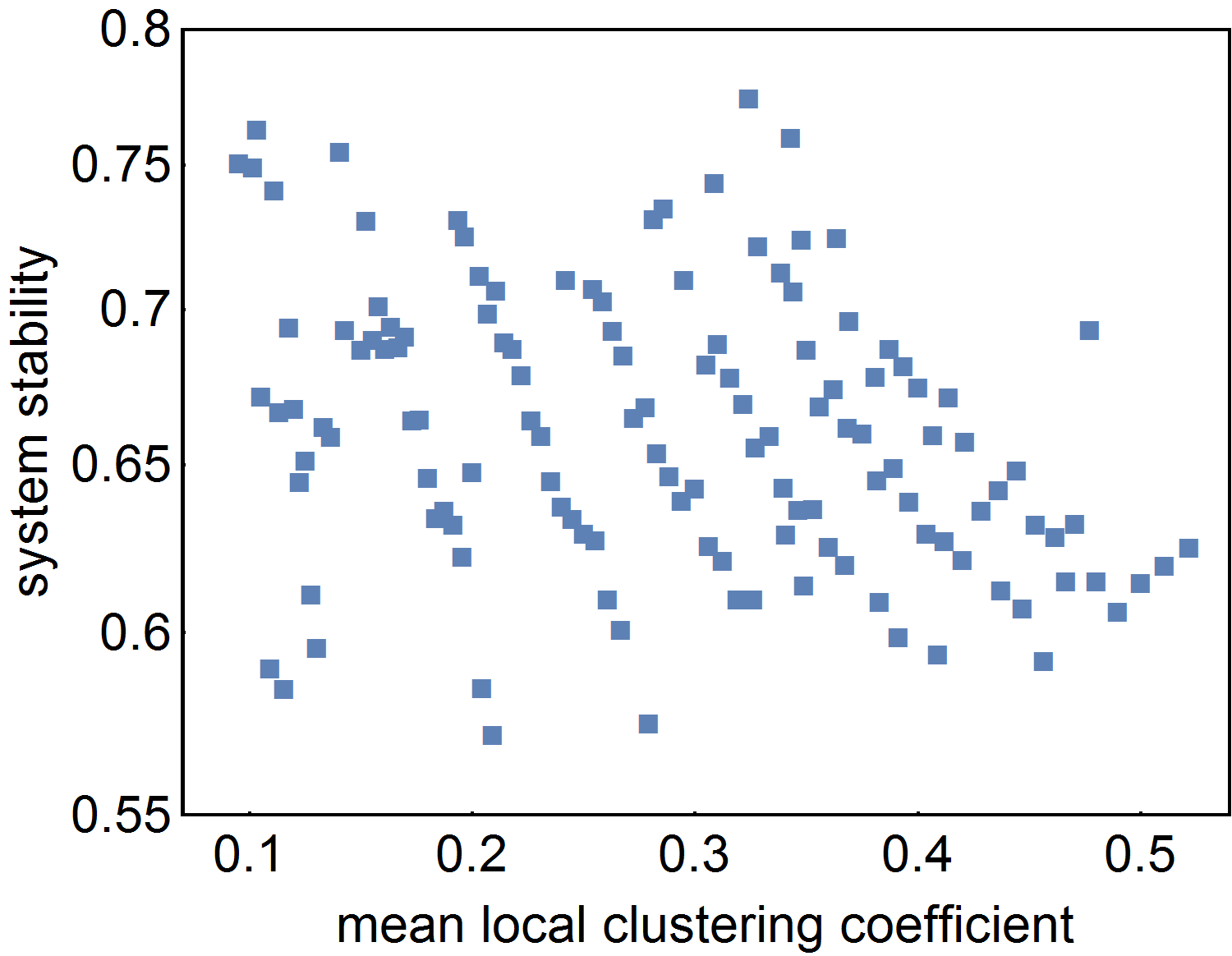}
\caption{Effect of the graph's mean local clustering on the stability of the $(R,G)$ system.}
\end{centering}
\end{figure}

It is important that we clarify how the results of this section can be interpreted. It is tempting to infer that graphs which underlie stable systems are themselves stable when subject to the addition and deletion of edges according to some consent rule.\footnote{Depending on the particular online community being studied, one might assume either a unilateral (Bala \& Goyal 2000) or mutual (Jackson \& Watts 2002) consent model for the alteration of the graph's edges. We disallow both here, but a future direction may involve easing this restriction to search for Nash equilibria in the action space of graphs for a given rate matrix.} However, recalling the formulation of the simulation, we tested the system's sensitivity with respect to perturbations of only the rate matrix, not the network itself. This does not invalidate the results; it merely highlights a necessary consequence of the implicit assumption that the matrix is fixed and agents have control over only their ``like'' rates. We must then understand the results of this section as describing which network features tend to support strategic rate ensembles. For an extreme example as to why this distinction matters, consider the star graph with $N=10$; that is, the graph with a single hub adjacent to each of the other nine nodes which are all mutually unconnected. Based on the positive correlation identified between the standard deviation of the degree distribution and the system's stability, it is reasonable to assume that stars have a propensity to support equilibrium ensembles. Indeed, it is found via simulation with the same parameters that 10-stars are on average $26.3\%$ more stable than the subset of Barab\'{a}si-Albert-generated networks with a hub of order 9 (which cannot contain stars due to the nature of the algorithm), which is already the most stable class identified (see Figure 3). However, among the $1\%$ most strategic star graphs, the branches have average likedness centrality of nearly three times that of the hub. The consequence for the system's stability is independent of the choice of Bala and Goyal's or Jackson and Watts' consent model. In either case, the hub has both incentive and permission to defect from the star, thus creating an isolated node with a likedness centrality of zero (by convention) and inflating its own prestige. We must conclude that even in this simple example, a system that is in Nash equilibrium over the action space of rate ensembles need not be stable when subjected to perturbations of the underlying network.

\section{Conclusion}

We return to our motivating question as to the incentive system driving the exchange of gratification in online social networks. The Facebook ``like'' and its counterparts elsewhere on the web, though desired and arbitrarily abundant, are traded as though they are scarce and may therefore permit an economic solution for the problem of their allocation. Here we studied a strictly strategic explanation which established likedness centrality as the effective utility function that social agents seek to maximize. We have seen that acting accordingly, systems of rational agents find equilibria at plausible finite rates of ``liking,'' indicating that strategic behavior may play a role in the individual's decision process. We acknowledge that one would be na\"{i}ve and excessively cynical to assert that this manner of Machiavellian reasoning is the only impetus behind the allocation of social gratification; it should be expected that the many other motivations identified in the literature and enumerated in the Introduction are also relevant. Nonetheless, the framework provided here may serve as the basis for a more exhaustive model of users' incentives. One might, for instance, consider a hybrid model akin to that of Jackson and Rogers (2007) whereby agents randomly decide at each iteration whether to maximize their likedness centrality or behave non-strategically. Then, carrying out the simulation for various probabilities, one can determine for a given application the relative prevalence of strategic versus non-strategic decision-making. New insights on social media use may emerge as it becomes possible to determine the degree to which particular societies and the individuals within them are driven by the pursuit of social capital.
\\ \\
The author is eternally grateful to Artur Filipowicz of Princeton University's Department of Operations Research and Financial Engineering for prioritizing revision and our useful discussions, even while they were neglected by our otherwise-engaged colleagues.

\section*{References}

\begin{hangparas}{.25in}{1}
Aigrain, P. (2012).\ \textit{Sharing: culture and the economy in the internet age.} Amsterdam University Press.

Albert, R., \& Barab\'{a}si, A. L. (2002).\ Statistical mechanics of complex networks. \textit{Reviews of modern physics, 74}(1), 47.

Appiah, K. A. (2011).\ \textit{The honor code: How moral revolutions happen.} WW Norton \& Company.

Arvidsson, A. (2009).\ The ethical economy: Towards a post-capitalist theory of value. \textit{Capital \& Class, 33}(1), 13-29.

Axelrod, R. M. (2006).\ \textit{The evolution of cooperation.} Basic books.

Bala, V., \& Goyal, S. (2000).\ A noncooperative model of network formation. \textit{Econometrica, 68}(5), 1181-1229.

Banerjee, A., Chandrasekhar, A. G., Duflo, E., \& Jackson, M. O. (2013).\ The diffusion of microfinance. \textit{Science, 341}(6144).

Barabási, A. L., \& Albert, R. (1999).\ Emergence of scaling in random networks. \textit{Science, 286}(5439), 509-512.

Bilton, N. (2014).\ Friends, and influence, for sale online. New York Times.

Brin, S., \& Page, L. (1998).\ The anatomy of a large-scale hypertextual Web search engine. \textit{Computer networks and ISDN systems, 30}(1), 107-117.

Castells, M. (2011).\ \textit{The rise of the network society: The information age: Economy, society, and culture (Vol. 1).} John Wiley \& Sons.

Cohen, B. (2003, June).\ Incentives build robustness in BitTorrent. In \textit{Workshop on Economics of Peer-to-Peer systems} (Vol. 6, pp.~68-72).

Curien, N., Muet, P. A., Cohen, E., Didier, M., \& Bordes, G. (2004).\ La société de l'information. \textit{La documentation française.}

Dehmer, M. (Ed.). (2011).\ \textit{Structural Analysis of Complex Networks.} Birkhäuser.

Ellison, N. B., Steinfield, C., \& Lampe, C. (2007).\ The benefits of Facebook ``friends:'' Social capital and college students' use of online social network sites. \textit{Journal of Computer-Mediated Communication, 12}(4), 1143-1168.

Fournier, L. (2014).\ Merchant Sharing Towards a Zero Marginal Cost Economy. \textit{arXiv preprint} arXiv:1405.2051.

Fuchs, C. (2009).\ Information and Communication Technologies and Society: A Contribution to the Critique of the Political Economy of the Internet. \textit{European Journal of Communication, 24}(1), 69-87.

Gerlitz, C., \& Helmond, A. (2011).\ Hit, Link, Like and Share. Organizing the social and the fabric of the web in a Like economy. In Paper presented at the \textit{DMI mini-conference} (Vol. 24, p.~25).

Jackson, M. O., \& Rogers, B. W. (2007).\ Meeting strangers and friends of friends: How random are social networks?. \textit{The American Economic Review}, 890-915.

Jackson, M. O., \& Watts, A. (2002).\ On the formation of interaction networks in social coordination games. \textit{Games and Economic Behavior, 41}(2), 265-291.

Jessen, J., \& J{\o}rgensen, A. H. (2011).\ Aggregated trustworthiness: Redefining online credibility through social validation. \textit{First Monday, 17}(1).

Joinson, A. N. (2008).\ Looking at, looking up or keeping up with people?: motives and use of facebook. In \textit{Proceedings of the SIGCHI conference on Human Factors in Computing Systems} (pp.~1027-1036). ACM.

Lerner, J., \& Tirole, J. (2002).\ Some simple economics of open source. \textit{The Journal of Industrial Economics, 50}(2), 197-234.

Raymond, E. S. (2001).\ \textit{The Cathedral \& the Bazaar: Musings on linux and open source by an accidental revolutionary.} O'Reilly Media, Inc.

Trivers, R. L. (1971).\ The evolution of reciprocal altruism. \textit{Quarterly Review of Biology}, 35-57.

Veale, K. (2003).\ Internet gift economies: Voluntary payment schemes as tangible reciprocity. \textit{First Monday, 8}(12).

Watts, D. J., \& Strogatz, S. H. (1998).\ Collective dynamics of ‘small-world' networks. \textit{Nature, 393}(6684), 440-442.

Weber, S. (2000).\ The political economy of open source software. \textit{Berkeley Roundtable on the International Economy.}
\end{hangparas}

\end{document}